\documentclass[preprint,12pt]{elsarticle}
\usepackage{a4wide,amssymb,amsmath,graphicx,color}

\biboptions{sort&compress,merge}
\journal{Physics Letters B}
\begin{document}

\begin{frontmatter}

\title{Neutrino spin oscillations in polarized matter}

\author[mipt,sinp]{A.~Grigoriev}
\ead{ax.grigoriev@mail.ru}
\author[mipt]{E.~Kupcheva}
\author[mipt]{A.~Ternov\corref{cor1}}
\ead{ternov.ai@mipt.ru}
\address[mipt]{Moscow Institute of Physics and Technology, 141701 Moscow Region, Dolgoprudny, Russia}
\address[sinp]{Skobeltsyn Institute of Nuclear Physics, Moscow State University, 119992 Moscow, Russia}
\cortext[cor1]{Corresponding author}


\begin{abstract}
We study the neutrino spin oscillations, i.e., neutrino spin precession caused by the neutrino interaction with matter polarized by external magnetic field (or, equivalently, by the interaction of the induced magnetic moment (IMM) of a neutrino with the magnetic field).
In the analysis, we consider realistic conditions inside supernovae and discuss both the Dirac and Majorana cases. We show that due to the interaction with the polarized matter a neutrino flux from a supernova suffers additional attenuation at low neutrino energies. Another possible effect is that when taken together the effects of conventional magnetic moment and polarized matter can cancel each other so that under certain condition the oscillations disappear. We note that this can lead to the appearance of a characteristic maximum in the spectrum of electron neutrinos from supernovae. Accounting for neutrino mixing (in the two-flavor approximation) can significantly increase (almost twice) the effective value of the IMM of the electron neutrino. As a result, electron neutrinos with twice as high energy will be able to participate in the processes of spin conversion. Various feedback mechanisms accompanying the considered phenomena are also discussed.
\end{abstract}

\end{frontmatter}


\section{Introduction \label{Intro}}

Investigation of electromagnetic properties of massive neutrinos, propagating in dense matter, is of great importance for elementary particle physics \cite{Giunti-Stud-RMP:2015}, as well as for neutrino astrophysics and cosmology \cite{Raffelt-Book-1996,Giunti-Kim-Book:2007,Dolgov-Cosmol:2002}.

It is of common knowledge that massive neutrinos can posses dipole magnetic moments $\mu_{ij}$ (diagonal for $i=j$ and transition for $i\neq j$; indices $i$ and $j$ denote neutrino mass states) \cite{Giunti-Stud-RMP:2015,Fuj-Shrock:80,Schecht-Valle:81bar,Nieves:82,Shrock:82,BorZhuKurTer:85e,Balantekin-Kayser:2018}.
Only Dirac neutrinos can have diagonal magnetic moments that lead to the spin oscillations phenomenon (i.e. helicity precession) in strong magnetic field \cite{Cisneros:71,Fuj-Shrock:80,Vol-Vys-Okun-JETF:86e}. Transition magnetic moments of massive Dirac and Majorana neutrinos interacting with magnetic field induce spin-flavor oscillations (the helicity precession accompanied by the change of neutrino flavor) \cite{Schecht-Valle:81bar,Akhmedov-PL-Main:88,*Lim-Marciano:88}.

In the framework of the minimally extended Standard Model (SM) with right-handed neutrino singlets added, the diagonal magnetic moment of the neutrino appeared to be very small \cite{Fuj-Shrock:80}:
\begin{equation}
\mu_{ii}=\mu=\frac{3eG_{\mathrm{F}}m}{8\sqrt{2}\pi^{2}}
\simeq3.2\times10^{-19}\mu_{\mathrm{B}}\,\left(  \frac{m}{1\,\mathrm{eV}}\right)  ,
\label{1-AMM-Classic}
\end{equation}
where $e$ is the absolute value of the electron charge, $m$ is the neutrino mass, $\mu_{\mathrm{B}}=e/2m_{e}$ is the Bohr magneton. The corresponding predictions for transition magnetic moments give even smaller values  \cite{Shrock:82,Raffelt-Book-1996,Giunti-Stud-RMP:2015}. Some extensions to the SM predict substantially larger values for the magnetic moments (for a review, see \cite{Giunti-Stud-RMP:2015}), and they do not contradict current experimental data. Experiments with reactor and solar neutrinos give bounds at a level $\mu\leq(2.8{-}2.9)\times10^{-11}\mu_{\mathrm{B}}$ \cite{GEMMA:2012,*Borexino-AMM:2017} (see also \cite{Balantekin-Vassh-AMM:2014,Canas-Valle:2016}), the analysis of astrophysical data lead to more stringent restriction: $\mu\leq(1.1{-}2.6)\times10^{-12}\mu_{\mathrm{B}}$
\cite{Viaux-clustM5-PRL:2013,*Arceo-Diaz-clust-omega:2015,*Kuznetsov:2009zm}. Values for magnetic moments close to the above mentioned  restrictions, turned out to be rather attractive for theoretical description of some astrophysical and cosmological phenomena \cite{Raffelt-Book-1996,Giunti-Stud-RMP:2015,Dolgov-Cosmol:2002,Balantekin-Kayser:2018}.

Nevertheless, there are certain reasons to consider the neutrino magnetic moments to be substantially smaller than the experimental bounds, even if one takes into account ``new physics'' beyond the minimally extended SM. In particular, in \cite{Bell-How-magnetic:2007} (see also references therein and \cite{Balantekin-AMM:2006,Balantekin-Kayser:2018}) on the basis of general model-independent analysis it is claimed that the Dirac neutrino magnetic moment must be bound at a level $\mu\leq3\times10^{-15}\mu_{\mathrm{B}}$. There exist also very stringent cosmological bounds (see \cite{Dolgov-Cosmol:2002} and references therein, and also \cite{Semikoz-Cosm-AMM:2017}).

The study of neutrino interactions with background medium enables revelation of various new mechanisms of neutrino spin oscillations, in which the helicity precession can take place without participation of the magnetic moment.  First of all, it should be mentioned that spin oscillations can be caused by the interaction of a neutrino with \emph{moving} medium, whose motion is transversal to the neutrino momentum. This effect has been predicted in \cite{Stud:2004e} (see also \cite{Lob-Stud:2001bar}), but has attracted  a considerable interest later (see, for instance,
\cite{Vlasenko-Full-Cirigl:2014,*Kart-Raffelt-Propag:2015,Dobr-Kart-Raffelt:2016}; a complete list of the relevant references can be found in
\cite{Studenikin-POS:2016,Dobr-Kart-Raffelt:2016,Pustoshny-Stud:2018}).

On the other hand, it is well known that when a neutrino moves in a dispersive medium the \emph{effective vertex} of electromagnetic neutrino interaction is modified
\cite{Or-Plah-Semikoz-Smor:87ee,Semikoz-Smor:89ee,Nieves-Pal-EM-neutr:89,*DOliv-Niev-Pal-EM-neutr:89,Altherr-Kainul-Ch-Med:91}.
Interaction of neutrinos with particles of background matter (for example, with
electrons, if a neutrino propagates in an electron plasma), leads to the appearance of new electromagnetic neutrino characteristics that
can take place only in the medium. One such characteristic is the \emph{induced neutrino magnetic moment}, which corresponds to the contribution of pseudovector currents of medium particles into the effective vertex of neutrino interaction
\cite{Semikoz-YAF:87ee,Semikoz-Smor:89ee,Or-Semikoz-Smor:94eee}.

The induced magnetic moment (IMM) can reach very high values in a medium. In particular, if the medium is a degenerate electron gas (the collapsing core of supernova, the interior of a neutron star), the IMM of the electron neutrino is given by \cite{Semikoz-YAF:87ee}
\begin{equation}
\mu^{\mathrm{ind}}=-\frac{eG_{\mathrm{F}}\mathtt{p}_{\mathrm{F}}}
{4\sqrt{2}\pi^{2}}\simeq-2.2\times10^{-13}\mu_{\mathrm{B}}\,\left(
\frac{\mathtt{p}_{\mathrm{F}}}{1\,\mathrm{MeV}}\right)  ,
\label{1-IMM-deg-el-gas}
\end{equation}
where $\mathtt{p}_{\mathrm{F}}$ is the Fermi momentum of the electron gas
\begin{equation}
\mathtt{p}_{\mathrm{F}}=\sqrt{\mu_{e}^{2}-m_{e}^{2}}\simeq130\times\left(
\frac{n_{e}}{10^{37}\,\mathrm{cm}^{-3}}\right)  ^{1/3}\mathrm{MeV},
\label{1-Imp-Fermi}
\end{equation}
$\mu_{e}$ and $n_{e}$ are the chemical potential and number density of electrons (see also \cite{Ternov-JETPL-2016ee,Ternov-PRD-2016} and references therein). For other neutrino flavors the sign of IMM is opposite to that of electron neutrino.

Formula (\ref{1-IMM-deg-el-gas}) refines the corresponding expression for IMM found earlier in \cite{Ternov-JETPL-2016ee,Ternov-PRD-2016}. (The expression (\ref{1-IMM-deg-el-gas}) differs from them by a common factor $-1/2$; this is confirmed by the independent verification\footnote{The IMM of the Dirac neutrino moving in a degenerate electron gas was independently calculated in Refs.
\cite{Eminov-Adv:2016,*Eminov-JETFe:2016} (and was called there as ``plasma correction to anomalous magnetic moment of neutrino''). The expression for IMM (provided that the corresponding contribution is correctly extracted from the radiative correction to neutrino energy) exactly reproduces the result (\ref{1-IMM-deg-el-gas}).}, and also by our calculations here, see Section~\ref{IMMonly}).

Note that the IMM of neutrino was studied in a considerable number of works (for example, \cite{Or-Plah-Semikoz-Smor:87ee,Semikoz-Smor:89ee,Nieves-Pal-EM-neutr:89,*DOliv-Niev-Pal-EM-neutr:89,Or-Semikoz-Smor:94eee}, a more complete list can be found in \cite{Dobr-Kart-Raffelt:2016,Ternov-PRD-2016}).
These investigations were mainly aimed at obtaining and analyzing the dispersion relations for neutrinos. Later the attention was drawn to the possibility of massive neutrino helicity change due to the interaction of the IMM with an external field \cite{Dobr-Kart-Raffelt:2016}. A detailed study of the massive Dirac neutrino spin evolution in this problem was carried out for the first time in Refs. \cite{Ternov-JETPL-2016ee,Ternov-PRD-2016}. The investigation was performed on the basis of the modified Dirac equation, taking into account the IMM contribution to the effective vertex of neutrino interaction with the external field. However in \cite{Ternov-JETPL-2016ee,Ternov-PRD-2016}, other types of neutrino interactions with background medium were not considered, in particular the effective neutrino potential in matter (the Mikheyev-Smirnov-Wolfenstein (MSW) potential \cite{Wolfen:78,*Mikh-Smirnov:85ee,Notz-Raffelt:88}).

In this letter, we further investigate neutrino spin oscillations caused by IMM, as well as by the simultaneous interaction of neutrino IMM and magnetic moment with an external magnetic field, taking consistently into account the effective neutrino potential in matter. Possible astrophysical consequences of the obtained results are analyzed. For the neutrino we consider both Dirac and Majorana cases and take matter to be homogeneous and at rest.

It should be noted that there exist another approach to physical interpretation of the phenomena described above. One may consider that the modification of the dispersion law, as well as the neutrino helicity evolution, occurs not due to the interaction of the IMM with external magnetic field, but rather due to neutrino scattering on medium particles polarized by the field (see, for example, \cite{Nunok-Semik-Smir-Val:97,Smirnov-Poland:97}, and also
\cite{Dobr-Kart-Raffelt:2016}). It is asserted that both points of view are equivalent and lead to the same results. As far as we know, a rigorous proof of this statement (for arbitrary neutrino energies and for arbitrary values of the magnetic field) is currently missing in the literature. Our paper can be considered as a proof of the above statement in the particular case of relativistic neutrinos ($E_{\nu}\gg m_{\nu}$) and relatively weak external magnetic field (see Eq.~(\ref{2-weak-field})) in a  medium at rest.

In this article, we will mainly follow the second approach (assuming that neutrinos interact with polarized particles of the medium), since it allows us to take consistently into account both the polarization of the medium and neutrino MSW potentials in matter.

\section{Spin oscillations of Dirac neutrino in polarized matter \label{IMMonly}}

First of all we consider spin oscillations of a Dirac neutrino caused by its interaction with a polarized medium, neglecting the contribution of the magnetic moment due to smallness of its value (\ref{1-AMM-Classic}) in the SM.

The additional energy, needed to describe spin oscillations of neutrinos coherently interacting with back\-ground particles, is obtained from the phenomenological Lagrangian that takes into account possible effects associated with matter motion and polarization \cite{Stud-Ternov-PLB:05bar,*Lobanov-SLnu:05bar} (see also \cite{Lob-Stud:2001bar,Grig-Lob-Stud:2002})
\begin{equation}
\Delta L_{\mathrm{eff}}=-f^{\mu}\left(  \bar{\nu}\,\gamma_{\mu}{\tfrac{1}%
{2}\left(  1+\gamma^{5}\right)  \,}\nu\right)  ,
\label{2-Lagrang}
\end{equation}
where $\gamma^{5}=-i\gamma^{0}\gamma^{1}\gamma^{2}\gamma^{3}$, and the 4-vector $f^{\mu}$ in the case of non-moving electrically neutral matter has the following components
\begin{equation}
f_{e}^{\mu}=\frac{G_{\mathrm{F}}}{\sqrt{2}}\left\{  2n_{e}-n_{n}
,\ -n_{e}\langle\boldsymbol{\zeta}_{e}\rangle\right\}  ,\quad f_{\mu}^{\mu
}=\frac{G_{\mathrm{F}}}{\sqrt{2}}\left\{  -n_{n},\ n_{e}\langle
\boldsymbol{\zeta}_{e}\rangle\right\}  \label{2-4vect-f}
\end{equation}
for electron and muon neutrino, respectively (the case of tau neutrino is analogous to that of muon one). Here $n_{e}$ and $n_{n}$ are
number densities of electrons and neutrons, $\langle\boldsymbol{\zeta}_{e}\rangle$ is the mean value for the polarization vector of the medium electrons \cite{Nunok-Semik-Smir-Val:97,Lob-Stud:2001bar,Grig-Lob-Stud:2002}.
It is assumed that electrons give the predominant contribution to the polarization of the medium, and neutrino mixing is not taken into account so that neutrino states with definite mass at the same time have definite flavor.

The external magnetic field is the source of the matter polarization\footnote{Various schemes of neutrino oscillations with account for the matter polarization were studied in \cite{Lob-Stud:2001bar,Grig-Lob-Stud:2002}. In these papers the polarization was not assumed to be caused by a magnetic field.}. In particular, the polarization of a degenerate electron gas arises from electrons occupying the ground Landau level, and in the weak-field limit
\begin{equation}
eB\ll\mathtt{p}_{\mathrm{F}}^{2} \label{2-weak-field}
\end{equation}
is linear in magnetic field $\mathbf{B}$ \cite{Nunok-Semik-Smir-Val:97}. Substituting into Eq.~(\ref{2-4vect-f}) the explicit expression for $\langle\boldsymbol{\zeta}_{e}\rangle$, found for these conditions in \cite{Nunok-Semik-Smir-Val:97}, we obtain
\begin{equation}
f_{e,\mu}^{\mu}=\left\{  \mathrm{V}_{e,\mu}\,,\ \mp2\mu^{\mathrm{ind}
}\mathbf{B}\right\}  ,\quad\mathrm{V}_{e}=\sqrt{2}G_{\mathrm{F}}\left(
n_{e}-n_{n}/2\right)  ,\quad\mathrm{V}_{\mu}=-\sqrt{2}G_{\mathrm{F}}\,n_{n}/2,
\label{2-4vect-f-last}
\end{equation}
where $\mu^{\mathrm{ind}}$ is given by Eq.~(\ref{1-IMM-deg-el-gas}). Without account for the potential $V_e$ in the expression for $f_{e}^{\mu}$, from the additional Lagrangian (\ref{2-Lagrang}) it is possible to recover the Dirac equation for the electron neutrino with IMM that completely coincides with the one used for description of the neutrino spin dynamics in
\cite{Ternov-JETPL-2016ee,Ternov-PRD-2016}.

Neutrino spin oscillations are typically described by the effective equation for the evolution of helicity amplitudes, which has the form of the Schr\"{o}dinger equation (see for instance \cite{Vol-Vys-Okun-JETF:86e})
\begin{equation}
i\frac{\partial}{\partial t}
\begin{pmatrix}
\nu_{e}^{s=-1}\\
\nu_{e}^{s=+1}%
\end{pmatrix}
=\left(  \mathrm{H}_{\mathrm{vac}}+\mathrm{H}_{\mathrm{matt}}+\mathrm{H}_{\mathrm{M}}\right)
\begin{pmatrix}
\nu_{e}^{s=-1}\\
\nu_{e}^{s=+1}
\end{pmatrix}
, \label{2-Schroed-like}
\end{equation}
where $s=\pm1$ is the neutrino helicity, $\mathrm{H}_{\mathrm{vac}}$ is the term of the effective Hamiltonian responsible for vacuum oscillations (in our case it is proportional to the identity matrix, and we will neglect it), $\mathrm{H}_{\mathrm{M}}$ is the term responsible for the interaction of magnetic moments with magnetic field (it will be considered in Section \ref{IMMandAMM}), $\mathrm{H}_{\mathrm{med}}$ is the contribution of matter, defined as $2{\times}2$-matrix with elements $\langle\nu_{e}^{s}|\Delta H_{\mathrm{eff}}|\nu_{e}^{s^{\prime}}\rangle$, where $\Delta H_{\mathrm{eff}}$ is the Hamiltonian, fol\-lowing from (\ref{2-Lagrang}).
To calculate the matrix elements, it is convenient to use the solutions of the free Dirac equation with an explicitly determined dependence on the neutrino helicity; these solutions can be found in Ref.~\cite{BorZhuTer:88ee}. As a result, we find
\begin{equation}
\mathrm{H}_{\mathrm{matt}}=
\begin{pmatrix}
\mathrm{V}_{e}  +2\mu^{\mathrm{ind}}B_{\Vert} & -\mu^{\mathrm{ind}}\gamma^{-1}B_{\bot}\\
-\mu^{\mathrm{ind}}\gamma^{-1}B_{\bot} & 0
\end{pmatrix}
, \label{2-Hmed}
\end{equation}
where $B_{\Vert}$ and $B_{\bot}$ are the longitudinal and transverse components of the magnetic field (with respect to the direction of neutrino motion), $\gamma=E_{\nu}/m_{\nu}$ is the neutrino gamma-factor.

The solution to equation (\ref{2-Schroed-like}) accounting for (\ref{2-Hmed}) for the case of constant matter density and magnetic field in terms of transition probability $\nu_{e}^{s=-1}\rightarrow\nu_{e}^{s=+1}$ at the time $t$ is
\begin{equation}
P(t)=\frac{(2\mu^{\mathrm{ind}}\gamma^{-1}B_{\bot})^{2}}
          {(\mathrm{V}_{e}+2\mu^{\mathrm{ind}}B_{\Vert})^{2}+(2\mu^{\mathrm{ind}}\gamma^{-1}B_{\bot})^{2}}
\sin^{2}\left\{  \sqrt{(\mathrm{V}_{e}+2\mu^{\mathrm{ind}}B_{\Vert})^{2}+(2\mu^{\mathrm{ind}}\gamma^{-1}B_{\bot})^{2}}\,\frac{t}{2}\right\}
\!. \label{2-Probab-IMM}
\end{equation}
In most of the real astrophysical conditions, for example, in a collapsing supernova, the probability (\ref{2-Probab-IMM}) turns out to be very small. This is because in (\ref{2-Probab-IMM}) the magnitude of the potential $\mathrm{V}_{e}$ typically is much larger than the terms containing $B$ due to the condition (\ref{2-weak-field}) \cite{Nunok-Semik-Smir-Val:97,Smirnov-Poland:97}.

However as it follows from (\ref{2-4vect-f-last}), the MSW potential of the electron neutrino $\mathrm{V}_{e}$ can turn to zero and in this case there can be realized conditions for the resonant transition of left-handed electron neutrinos produced in the supernova core, into sterile right-handed states that practically do not interact with matter.
Indeed, the expression for $\mathrm{V}_{e}$ in (\ref{2-4vect-f-last}) can be presented in the form \cite{Vol-Vys-Okun-JETF:86e,Voloshin-SN-JETF:88e,*Voloshin-SN-PLB:88}
\begin{equation}
\mathrm{V}_{e}=\sqrt{2}G_{\mathrm{F}}\frac{\rho_{\mathrm{B}}}{m_{\mathrm{N}}
}\left(  Y_{e}-Y_{n}/2\right)  =\frac{G_{\mathrm{F}}}{\sqrt{2}}\frac
{\rho_{\mathrm{B}}}{m_{\mathrm{N}}}\left(  3Y_{e}-1\right)  , \label{2-Ve}
\end{equation}
where $\rho_{\mathrm{B}}$ is the mass density and $\rho_{\mathrm{B}}/m_{\mathrm{N}}=n_{\mathrm{B}}=n_{n}+n_{p}$ is the number density of  nucleons, $Y_{e}=n_{e}/n_{\mathrm{B}}=n_{p}/n_{\mathrm{B}}$, $Y_{n}=n_{n}/n_{\mathrm{B}}$ is the number of the electrons, protons and neutrons per baryon, respectively. The resonance condition for neutrino spin oscillations following from (\ref{2-Ve}) reads as
\begin{equation}
Y_{e}\approx1/3. \label{2-Rez}
\end{equation}

Condition (\ref{2-Rez}) can be realized in the envelope of the collapsing supernova. The neut\-ronization pulse, during which the star emits electron neutrinos only \cite{Mirizzi-Tamborra:2016} (lasting the first $\sim 25$~ms after the core bounce), is accompanied by a substantial deleptonization of the stellar matter. As a result, a characteristic dip occurs in the radial dependence $Y_{e}(r)$ ($r$ is the distance from the center of the star) when $Y_{e}$ first falls down to $\sim 0.1$, and then increases to $Y_{e}\sim0.5$ \cite{Janka-Lang:2007,*Burrows-RMP:2013}. Therefore, there must exist a point (at $dY_{e}/dr>0$), in which the resonance condition (\ref{2-Rez}) is satisfied (see also \cite{Anik-Kuz-Mikh-AMM-SN:2010e}). We note that these features of the dependence $Y_{e}(r)$ are retained for a time longer than the duration of the neutronization pulse \cite{Janka-Lang:2007,*Burrows-RMP:2013}. Thus the condition (\ref{2-Rez}) can also be satisfied for electron neutrinos (and antineutrinos) emitted during the subsequent phases of neutrino emission by the star\footnote{The resonance condition (\ref{2-Rez}) can also be realized in compact object mergers (see, for example, \cite{Wanajo:2014,Sumiyoshi-NS:2017}), as well as in models involving sterile neutrinos \cite{Nunokawa-Peltoniemi-ster:97,Wu-Fischer-ster:2014}.}.

The resonance condition in a collapsing supernova (\ref{2-Rez}) corresponds to density $\rho_{\mathrm{B}}\sim 10^{9}{-}10^{12}\ \text{g/cm}^{3}$ above the neutrinosphere\footnote{The effective potential (\ref{2-Ve}) could also include a term corresponding to neutrino interaction with neutrinos of the medium ($\nu_{e} - \nu_{e} $ scattering). This could change the resonance condition (\ref{2-Rez}), see Section~\ref{Feedbacks}. In our case, this term can be neglected, since in the considered region the neutrino density is much lower than the electron one \cite{Nunokawa-Peltoniemi-ster:97,Wu-Fischer-ster:2014}. See, however, the discussion in \cite{Blinn-Okun:88e,*Peltoniemi:92,*Yudin-Fomichev:2016e}.}
\cite{Nunokawa-Peltoniemi-ster:97,Wu-Fischer-ster:2014,Sumiyoshi-NS:2017,Anik-Kuz-Mikh-AMM-SN:2010e}. For effective conversion of neutrinos at the resonance point, the adiabatic condition must be satisfied, requiring that the width of the resonance is of the order of or greater than half of the oscillation length \cite{Athar-Peltoniemi-2:95,*Athar-Peltoniemi:95,Dighe-Smirnov:2000}. In our case this lead to condition
\begin{equation}
\varkappa=\frac{2(2\mu_{\nu}^{\mathrm{ind}}\gamma^{-1}B_{\bot})^{2}%
}{\left\vert d\mathrm{V}_{e}/dr\right\vert }\gtrsim1. \label{2-Ad-gen}
\end{equation}
If the adiabaticity parameter $\varkappa\gg1$, then  a complete conversion of the left-handed neutrinos into the right-handed takes place. When the adiabaticity condition is violated the survival probability can be estimated by the Landau-Zener formula \cite{Landau:1932,*Zener:1932,Dighe-Smirnov:2000} (see also \cite{Haxton:86})
\begin{equation}
P_{\nu_{e}^{s=-1}\rightarrow\nu_{e}^{s=-1}}=\exp\left(  -\frac{\pi}
{2}\varkappa\right)  . \label{2-LandZen}
\end{equation}

Condition (\ref{2-Ad-gen}) leads to the following restriction on the strength of the magnetic field:
\begin{equation}
B\gtrsim1.2\times10^{12}\gamma\left(  \frac{\rho_{B}}{10^{12}~\text{g/cm}^{3}}\right)^{1/6}\left(  \frac{dY_{e}/dr}{10^{-8}~\text{cm}^{-1}}\right)^{1/2}\left(  \frac{Y_{e}}{1/3}\right)^{-1/3}\text{G}. \label{2-Ad-field}
\end{equation}
Conditions (\ref{2-Ad-gen}) and (\ref{2-Ad-field}) depend on the neutrino gamma factor $\gamma=E_{\nu}/m_{\nu}$, and this is an interesting feature of the phenomenon under consideration. If we fix the value of the field strength $B$, then neutrinos with lower energies will be more efficiently converted into right-handed than neutrinos with higher energies. On the other hand, the greater the neutrino energy, the stronger the field is required to satisfy the adiabaticity condition (\ref{2-Ad-field}). In addition, condition (\ref{2-Ad-field}) turns out to be very sensitive with respect to the value of $dY_{e}/dr$. That is, for effective neutrino conversion, a smooth dependence $Y_{e}(r)$ is needed.

The field strength is also bounded from above, since we use the weak-field approximation (\ref{2-weak-field}). From Eq.~(\ref{2-weak-field}) with account for (\ref{1-Imp-Fermi}) we obtain
\begin{equation}
B\ll2.2\times10^{17}\left(  \frac{\rho_{B}}{10^{12}~\text{g/cm}^{3}}\right)
^{2/3}\left(  \frac{Y_{e}}{1/3}\right)  ^{2/3}\text{G}.
\label{2-weak-field-2}%
\end{equation}

\begin{figure}[t!]
 \centering
      \begin{minipage}{0.49\hsize}
        \begin{center}
          \includegraphics[width=0.95\linewidth]{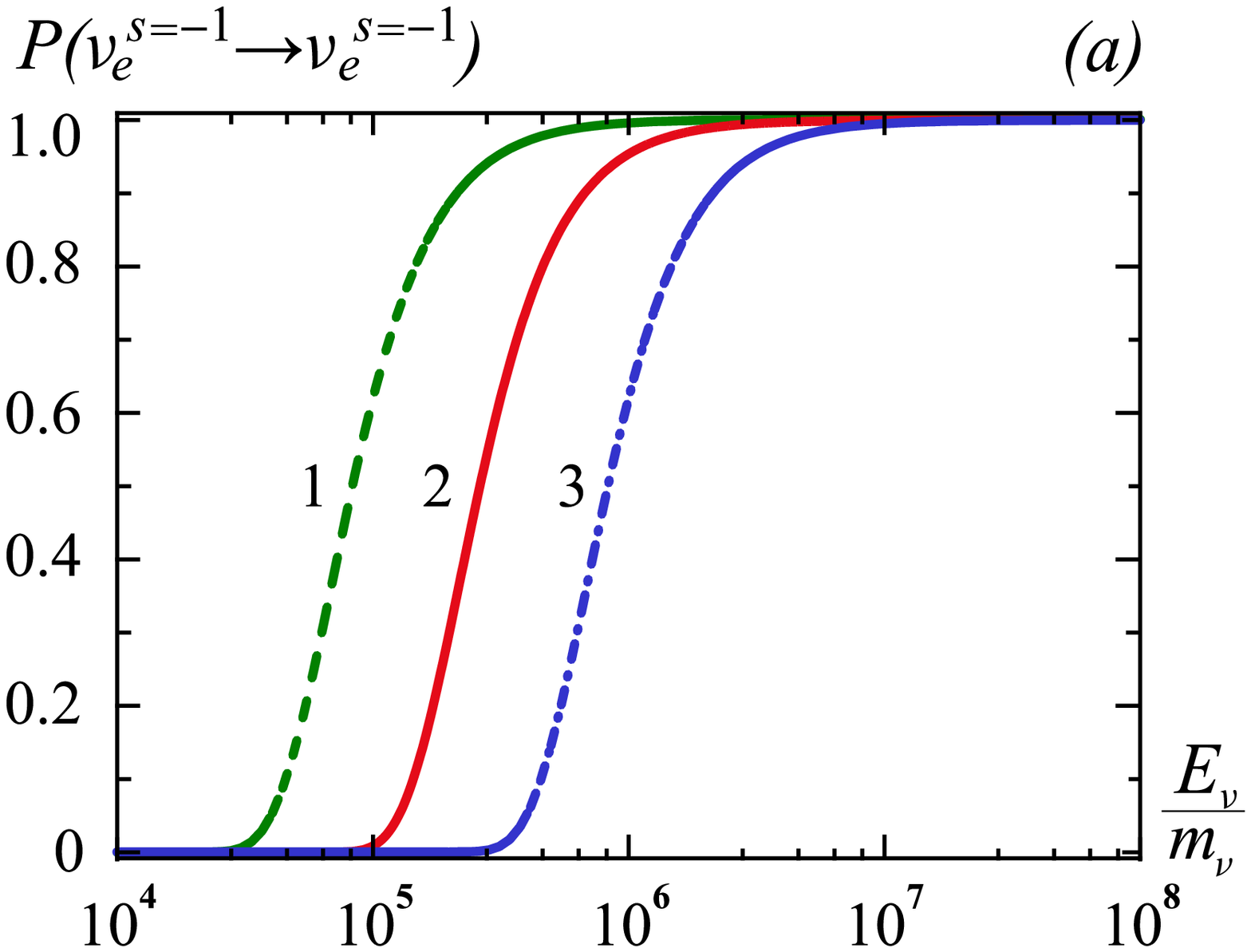}
        \end{center}
      \end{minipage}
      \hfill
      \begin{minipage}{0.49\hsize}
        \begin{center}
          \includegraphics[width=0.95\linewidth]{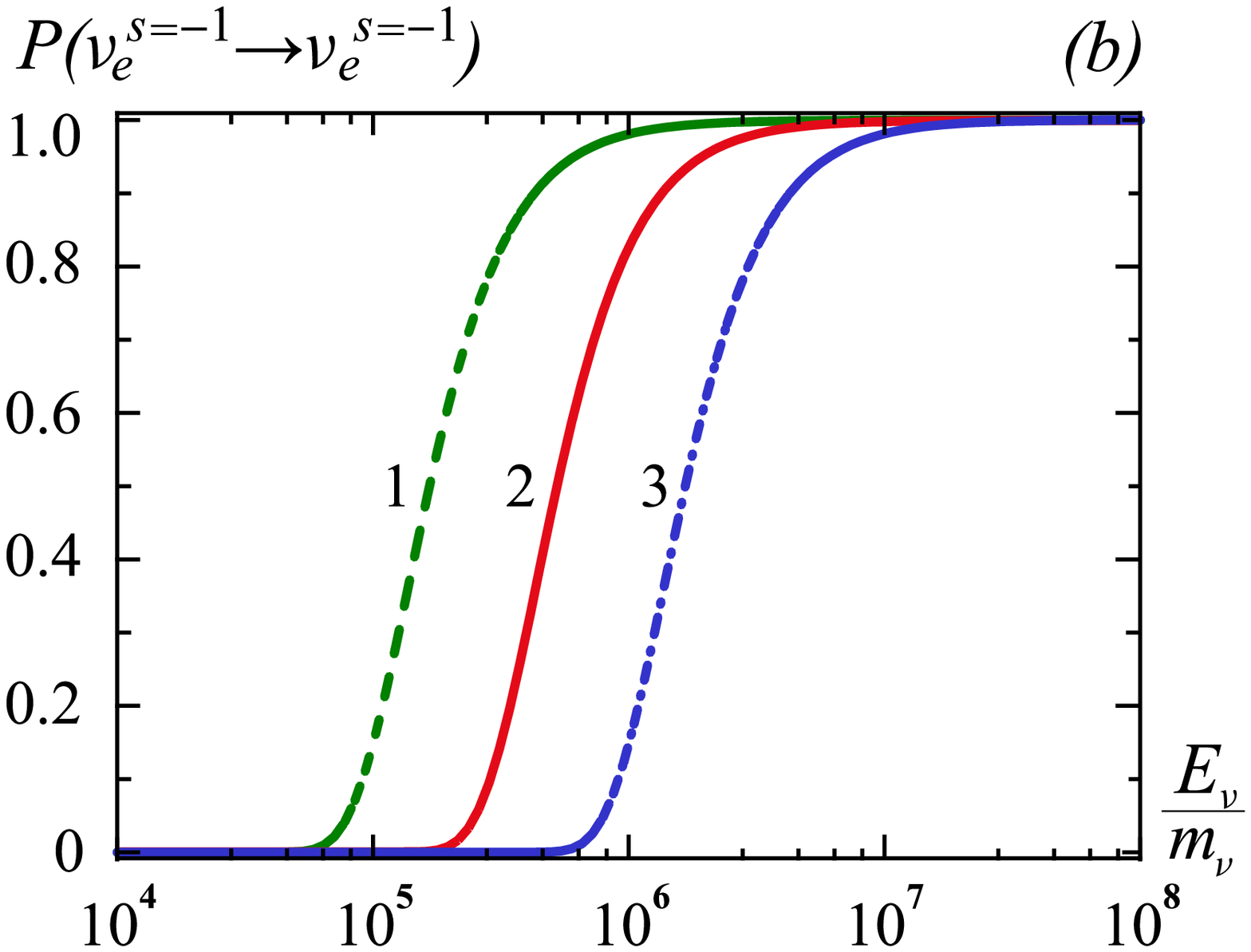}
        \end{center}
      \end{minipage}
\caption{The survival probability for neutrino with negative helicity as a function of neutrino energy: (a) -- for Dirac neutrino, (b) -- for Majorana neutrino; Line~1: $dY_{e}/dr=10^{-8}\ $cm$^{-1}$, Line~2: $dY_{e}/dr=10^{-9}\ $cm$^{-1}$, Line~3: $dY_{e}/dr=10^{-10}\ $cm$^{-1}$; $\rho_{\mathrm{B}}=10^{12}\ $g/cm$^{3}$, $B=6.6\times10^{16}$ G, $Y_{e}=1/3$.  \label{Fig1}}
\end{figure}
Fig.~\ref{Fig1}a shows the probability of the resonant transition (the survival probability $\nu_{e}^{s = -1}\rightarrow\nu_ {e}^{s = -1} $) depending on the neutrino energy and on the value of the electron fraction gradient $dY_{e}/dr$. It can be seen that there is a region of relatively low energies where neutrinos undergo conversion from left- to right-handed almost completely, followed by a region of energies where neutrinos are only partially converted. Finally, at high energies, the neutrino conversion is almost absent, that is, the so-called strong adiabaticity violation is realized~\cite{Dighe-Smirnov:2000}.

The energy range where the conversion takes place is determined by the value of $dY_{e}/dr$. For example, for $dY_{e}/dr\simeq10^{-9}\ $cm$^{-1}$ (the curve 2 in Fig.~\ref{Fig1}a) the conversion affects neutrinos with energies up to $1$ MeV (for $m_{\nu}\simeq1$ eV \cite{PDG:2016bar}). At that, neutrinos with energies up to $0.1$ MeV are converted completely. For $dY_{e}/dr\simeq10^{-8}\ $cm$^{-1}$ the overall picture of the transitions is preserved, but the energy boundary at which the conversion becomes effective shifts towards lower energies (see Fig.~\ref{Fig1}a).

The derivative $dY_{e}/dr$ depends on the time elapsed since the core bounce and behaves differently in different models of collapse. First, we note that the value of $dY_{e}/dr\simeq10^{-9}\ $cm$^{-1}$ is characteristic for small time intervals ($t < 3$ ms) following the bounce \cite{Kuroda-SN:2016} (thus our effect will lead to suppression of the low-energy part of the spectrum of the initial neutrino flux). Then, there appears the above mentioned  dip in the $Y_{e}(r)$ dependence. Further, the derivative $dY_{e}/dr$ decreases at the point $Y_{e}=1/3$ for some time, and by ${\simeq}{100{-}150}$~ms after the bounce reaches a level of ${\sim}10^{-9}{-}10^{-8}\ $cm$^{-1}$ \cite{Nunokawa-Peltoniemi-ster:97,Pan-Liebendorfer-SN:2016,Kuroda-SN:2016,Janka-NS:2018}. After~that $dY_{e}/dr$ grows fast, and by ${\sim}500$~ms the effect of neutrino flux attenuation should disappear.

By this means, we predict the attenuation of the low-energy part of the neutrino signal from the supernova for some limited time after collapse. We emphasize that so far we have not taken into account the presence of the neutrino magnetic moment, assuming that it is small (\ref{1-AMM-Classic}) and does not affect the neutrino spin dynamics.

We note that for the estimates above, we have used the value of the field strength equal to $B=6.6\times10^{16}$~G, that satisfy the condition (\ref{2-weak-field-2}). This value does not contradict current research, according to which the generated in supernova envelope fields can reach magnitudes ${\sim}10^{16}{-}10^{17}$~G \cite{Ardeljan:2005,*Burrows-rot:2007,*Potekhin-field:2015}.

\section{Spin oscillations of Dirac neutrino
with the magnetic moment \\ in polarized matter \label{IMMandAMM}}

In this section, we consider spin oscillations of Dirac neutrinos, caused not only by the interaction with polarized medium, but also by the ``direct'' interaction of the neutrino magnetic moment with the magnetic field. Spin oscillations caused by the magnetic moment (without taking into account the matter polarization) have been considered in a number of papers, see, for instance\footnote{In papers \cite{Arb-Lobanov-Mur:2009e,*Arb-Lobanov-Mur:2010}, the spin dynamics of a Dirac neutrino with a magnetic moment in an external field and in a medium is considered, but the neutrino helicity is determined by a method different from ours.}, \cite{Cisneros:71,Fuj-Shrock:80,Vol-Vys-Okun-JETF:86e,Voloshin-SN-JETF:88e,*Voloshin-SN-PLB:88,Likh-Stud-JETP:95e,Anik-Kuz-Mikh-AMM-SN:2010e,Lychk-Blinn-SN-AMM:2010}
(a more complete list can be found in \cite{Giunti-Stud-RMP:2015}). The corresponding contribution to the effective Hamiltonian in Eq.~(\ref{2-Schroed-like}) has the form \cite{Giunti-Stud-RMP:2015,Fabbr-Grig-Stud:2016,Akhmedov-Long-MPhLett:88}
\begin{equation}
\mathrm{H}_{\mathrm{M}}=
\begin{pmatrix}
\mu_{\nu}\gamma^{-1}B_{\Vert} & -\mu_{\nu}B_{\bot}\\
-\mu_{\nu}B_{\bot} & -\mu_{\nu}\gamma^{-1}B_{\Vert}
\end{pmatrix}
, \label{4-H-AMM}
\end{equation}
where $\mu_{\nu}$ is the diagonal neutrino magnetic moment.  Using (\ref{4-H-AMM}) the evolution equation (\ref{2-Schroed-like}) leads to the following expression for the probability of the transition $\nu_{e}^{s=-1}\rightarrow\nu_{e}^{s=+1}$ at time $t$ (to be compared with Eq.~(\ref{2-Probab-IMM})):
\begin{equation}
P(t)=\frac{\left(  2(\mu_{\nu}^{\mathrm{ind}}\gamma^{-1}+\mu_{\nu})B_{\bot
}\right)  ^{2}}{\left(  2(\mu_{\nu}^{\mathrm{ind}}\gamma^{-1}+\mu_{\nu
})B_{\bot}\right)  ^{2}+\left(  \mathrm{V}_{e}+2(\mu_{\nu}^{\mathrm{ind}}%
+\mu_{\nu}\gamma^{-1})B_{\Vert}\right)  ^{2}}\sin^{2}\left\{  \sqrt{D}%
\,\frac{t}{2}\right\}  , \label{4-Prob-IMM-AMM}
\end{equation}
where $D$ is the denominator of the pre-sine factor.

The most important feature of expression (\ref{4-Prob-IMM-AMM}), which is manifested when the magnetic moment and the medium polarization influence the neutrino spin dynamics simultaneously, is the possibility for disappearance of oscillations  under the condition
\begin{equation}
\mu_{\nu}^{\mathrm{ind}}\gamma^{-1}=-\mu_{\nu}. \label{4-Anti-Rez}
\end{equation}
Equality (\ref{4-Anti-Rez}) can be realized if the quantities $\mu_{\nu}^{\mathrm{ind}}$ and $\mu_{\nu}$ have different signs. This is precisely the case for the electron neutrino in minimally extended SM, see (\ref{1-AMM-Classic}) and (\ref{1-IMM-deg-el-gas}). Indeed, from (\ref{4-Prob-IMM-AMM}) it follows that at the resonance point $Y_{e}\approx1/3 $ (neglecting, as before, the longitudinal field effects) and when (\ref{4-Anti-Rez}) is fulfilled, the frequency of spin oscillations $\sqrt{D}\rightarrow0$.

\begin{figure}[t!]
      \centering{\includegraphics[width=0.48\linewidth]{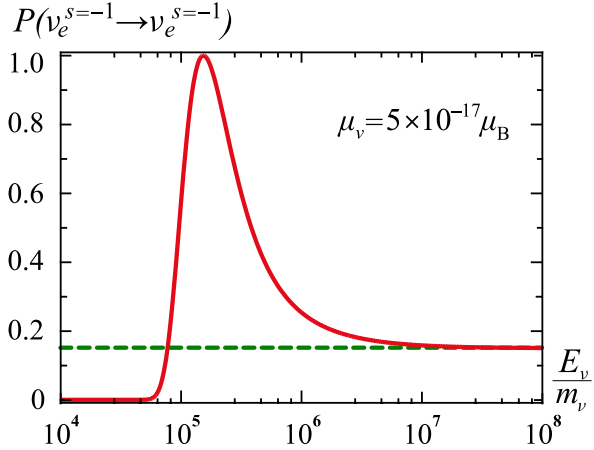}}
\caption{The survival probability for Dirac neutrino with magnetic moment and with negative helicity depending on the neutrino energy: $dY_{e}/dr=10^{-9}\ $cm$^{-1}$, $\rho_{\mathrm{B}}=10^{12}\ $g/cm$^{3}$, $B=6.6\times10^{16}$ G, $Y_{e}=1/3$, $\mu_{\nu}=5\times10^{-17}\mu_{\mathrm{B}}$. The dashed line corresponds to the conversion due to the magnetic moment only. \label{Fig2}}
\end{figure}
Note that the effect of ceasing the neutrino helicity conversion in an external magnetic field on the condition (\ref{4-Anti-Rez}) was first predicted in \cite{Ternov-PRD-2016}.

The main features of the resonant transitions of the left-handed electron neutrinos into sterile right-handed states in the collapsing supernova can be understood from the analysis of the adiabaticity condition, which now should read as
\begin{equation}
\varkappa_{\mathrm{M}}=\frac{2(2(\mu_{\nu}^{\mathrm{ind}}\gamma^{-1}+\mu_{\nu
})B_{\bot})^{2}}{\left\vert d\mathrm{V}_{e}/dr\right\vert }\gtrsim1,
\label{4-Ad-AMM}
\end{equation}
and from Eq.~(\ref{2-LandZen}). Since we are in the same framework of basic assumptions concerning the medium and the magnetic field as in Section \ref{IMMonly}, the resonance condition for spin oscillations (\ref{2-Rez}) does not change. On the basis of the data used in Section~\ref{IMMonly} we have $\mu_{\nu}^{\mathrm{ind}}=-7.9\times10^{-12}\mu_{\mathrm{B}}$, and let us for definiteness set $\mu_{\nu}=5\times10^{-17}\mu_{\mathrm{B}}$.

If the neutrino energy is not very high i.e. $\left\vert \mu_{\nu}^{\mathrm{ind}}\gamma^{-1}\right\vert \gg\left\vert \mu_{\nu}\right\vert$, then the effect of matter polarization on the conversion process at the resonance will be determining (see Fig.~\ref{Fig2}), and the survival probability $\nu_{e}^{s=-1}\rightarrow\nu_{e}^{s=-1}$ behaves the same way as in the absence of a magnetic moment (Fig.~\ref{Fig1}a). Further, with increasing neutrino energy, the condition $\left\vert \mu_{\nu}^{\mathrm{ind}}\gamma^{-1}\right\vert \sim\left\vert\mu_{\nu}\right\vert $ becomes valid. In this case, the survival probability rises sharply to the value $P_{\nu_{e}^{s=-1}\rightarrow\nu_{e}^{s=-1}}=1$,  so that there are no transitions with change in the helicity. In our case (Fig.~\ref{Fig2}), the maximum of the probability corresponds to the energy ${\sim}0.16$ MeV; for a larger value $\mu_{\nu}$, the maximum is achieved at lower neutrino energies. And finally, at $\left\vert \mu_{\nu}\right\vert \gg\left\vert \mu_{\nu}^{\mathrm{ind}}\gamma^{-1}\right\vert $, the magnetic moment will have a determining influence on the conversion process. The degree of attenuation of the neutrino flux will depend on the adiabaticity condition for the transitions that are due to the magnetic moment only \cite{Anik-Kuz-Mikh-AMM-SN:2010e}. As can be seen from (\ref{4-Ad-AMM}), it does not depend on energy, so in this case the conversion will affect neutrinos of all energies.

It should be noted that observations of the characteristic maximum in the spectrum of electron neutrinos from supernova (see Fig.~\ref{Fig2}) at relatively low energies $E\lesssim1$~MeV, may indicate that neutrinos can possess a sufficiently large magnetic moment (for our conditions, at least $\mu_{\nu}\gtrsim 10^{-17}\mu_{\mathrm{B}}$).

\section{Spin oscillations of Majorana neutrino in polarized matter \label{IMMonlyMN}}

As it is known, the Majorana neutrino, being identical to its antiparticle, can not possess a diagonal magnetic moment in the vacuum (see, for example, \cite{Giunti-Stud-RMP:2015} and literature cited therein). The IMM of such a neutrino in a medium, however, can be non-zero if there is no symmetry between background particles and antiparticles \cite{Nieves-Pal-EM-neutr:89,*DOliv-Niev-Pal-EM-neutr:89,Semikoz-Smor:89ee,Dobr-Kart-Raffelt:2016}. Consequently, spin oscillations, caused by the interaction of the Majorana neutrino with a polarized medium, can take place. The difference from the Dirac case is that left-handed neutrinos transit to right-handed states, which are no longer sterile. According to the existing terminology, it is customary to call Majorana neutrinos with right helicity as ``antineutrinos'', since their weak interactions are the same as those of the Dirac antineutrinos \cite{Giunti-Stud-RMP:2015}.

The effective Hamiltonian for Eq.~(\ref{2-Schroed-like}) that describes the interaction of Majorana neutrinos with matter accounting for polarization of the electrons  can be obtained by an analogy with (\ref{2-Hmed}) on the basis of the effective Lagrangian for Majorana neutrino $\Delta L_{\mathrm{eff}}$
\cite{Grig-Stu-Ternov-Major:2006}:
\begin{equation}
\mathrm{H}_{\mathrm{med}}^{M}=%
\begin{pmatrix}
\mathrm{V}_{e}+2\mu_{\nu}^{\mathrm{ind}}B_{\Vert} & -2\mu_{\nu}^{\mathrm{ind}%
}\gamma^{-1}B_{\bot}\\
-2\mu_{\nu}^{\mathrm{ind}}\gamma^{-1}B_{\bot} & -\mathrm{V}_{e}-2\mu_{\nu
}^{\mathrm{ind}}B_{\Vert}
\end{pmatrix}
, \label{3-Hmed-Maj}
\end{equation}
where the notations coincide with those used in (\ref{2-Hmed}). Further, performing calculations analogous to ones in section \ref{IMMonly}, one can obtain an expression for the probability of the transition $\nu_{e}^{s=-1}\rightarrow\nu_{e}^{s=+1}$ in a magnetic field where the amplitude factor will coincide with the corresponding factor in (\ref{2-Probab-IMM}), and the oscillation frequency is doubled in comparison with (\ref{2-Probab-IMM}).

Applying our results to description of the resonance transitions $\nu_{e}^{s=-1}\rightarrow\nu_{e}^{s=+1}$ (i.e. $\nu_{e}\rightarrow\bar{\nu}_{e}$) in the conditions of the collapsing supernova (see section~\ref{IMMonly}) we first of all  observe that the resonance condition (\ref{2-Rez}) does not change. The adiabaticity condition (\ref{2-Ad-gen}) must be modified, the parameter $\varkappa$ in the case of Majorana neutrinos doubles: $\varkappa\rightarrow2\varkappa$. This leads to the fact that in the conversion $\nu_{e}\rightarrow\bar{\nu}_{e}$ involved will be  higher-energy neutrinos  in comparison with the Dirac case (with energies up to ${\sim}2$~MeV under the same conditions as in the section~\ref{IMMonly}, see Fig.~\ref{Fig1}b). Other conclusions of section~\ref{IMMonly} hold for the case of the Majorana neutrino, with the difference that now we should speak not about attenuation, but about \emph{distortion} of the low-energy part of the neutrino signal from the supernova associated with the resonance conversion $\nu_{e}\leftrightarrows\bar{\nu}_{e}$.

\section{Effects of neutrino mixing\label{Mixing}}

Now we discuss the impact of neutrino mixing on the phenomena under
consideration. It is known that in general the mixing can reduce the value of
neutrino electromagnetic moments
\cite{Balantekin-Vassh-AMM:2014,Canas-Valle:2016}.

Let us consider in more detail the situation concerning the IMM, using the two-flavor approximation. First, we transit into the flavor basis using the relations
\begin{equation}
\nu_{e}^{\pm}=\nu_{1}^{\pm}\cos\theta+\nu_{2}^{\pm}\sin\theta,\quad\nu_{\mu
}^{\pm}=-\nu_{1}^{\pm}\sin\theta+\nu_{2}^{\pm}\cos\theta, \label{5-Fl-basis}%
\end{equation}
where the sign ``$\pm$'' denotes the helicity of the neutrino state. Exactly these states should be used in calculating the effective evolution Hamiltonian (which is now a $4{\times}4$-matrix). As a result we obtain that for the electron neutrino the equation governing transitions with change of  helicity coincides with (\ref{2-Schroed-like}), but the Hamiltonian $\mathrm{H}_{\mathrm{matt}}$ now has the form
\begin{equation}
\mathrm{H}_{\mathrm{matt}}^{e}=
\begin{pmatrix}
\mathrm{V}_{e}+2\mu_{ee}^{\mathrm{ind}}B_{\Vert} & -\mu_{ee}^{\mathrm{ind}%
}\gamma^{-1}B_{\bot}\\
-\mu_{ee}^{\mathrm{ind}}\gamma^{-1}B_{\bot} & 0
\end{pmatrix}
, \label{5-Hmed-mix}
\end{equation}
where $\mu_{ee}^{\mathrm{ind}}=\mu^{\mathrm{ind}}(1+\sin2\theta)$ is the IMM of the electron neutrino in the flavor basis\footnote{Note that there are no transition induced moments of the type $\mu_{e\mu}^{\mathrm{ind}}$, unless the nonstandard neutrino interactions (NSI) are considered \cite{Ohlsson-NSI:2013,*Miranda-Nunokawa-NSI:2015}.}, $\mu^{\mathrm{ind}}$ is given by the usual formula (\ref{1-IMM-deg-el-gas}). In equation (\ref{5-Hmed-mix}) we have adopted the following approximation $\gamma^{-1}\approx\gamma_{1}^{-1}\approx\gamma _{2}^{-1}\approx\gamma_{12}^{-1}$, where $\gamma_{1,2}^{-1}=m_{1,2}/E_{1,2}$ and $\gamma_{12}^{-1}=(\gamma_{1}^{-1}+\gamma_{2}^{-1})/2$. By this means, $\gamma$ is the common neutrino $\gamma$-factor. It is clear that, using the method described above, one can also obtain the generalization of the effective Hamiltonian (\ref{3-Hmed-Maj}) for Majorana neutrino.

One can see that the modification of the Hamiltonian describing the interaction of the left-handed neutrino with a polarized medium
(\ref{5-Hmed-mix}) leads to effective shift of the IMM, $\mu^{\mathrm{ind}}\rightarrow\mu_{ee}^{\mathrm{ind}}=\mu^{\mathrm{ind}}(1+\sin2\theta)$.
Therefore, using the solar neutrino mixing angle (according to the latest data \cite{deSalas-Fit:2018,*Esteban-Fit:2019}, $\theta_{12}\simeq34.5^{\circ}$, hence $\sin2\theta\simeq0.93$), we have an almost twofold increase in the IMM value.

This leads to the following changes in description of the phenomena presented above.

First, we have an increase in adiabaticity for neutrino transitions.
In the expressions for adiabaticity parameters (\ref{2-Ad-gen}) and (\ref{4-Ad-AMM}), it is necessary to make a replacement $\mu^{\mathrm{ind}}\rightarrow\mu_{ee}^{\mathrm{ind}}$. This means that neutrinos with even higher energies (up to ${\sim}2$ MeV for Dirac and up to ${\sim}4$ MeV for Majorana neutrinos under the conditions discussed in sections~\ref{IMMonly} and~\ref{IMMonlyMN}) will participate in transitions with a helicity flip. At the same time, the condition (\ref{2-Ad-field}) is softened, so that the range of the $B$ field where the adiabaticity holds becomes twice as wide due to its lower boundary reduction.

Second, the condition for disappearance of Dirac neutrino spin oscillations (\ref{4-Anti-Rez}) due to competition between the IMM and the diagonal magnetic moment now takes the form
\begin{equation}
\mu_{ee}^{\mathrm{ind}}\gamma^{-1}=-\mu_{ee}, \label{5-Anti-Rez-Flav}%
\end{equation}
where
\[
\mu_{ee}=\mu_{11}\cos^{2}\theta+\mu_{22}\sin^{2}\theta+\mu_{12}\sin2\theta
\]
is the effective diagonal magnetic moment of the electron neutrino in the flavor basis (see, for example, \cite{Pustoshny-Stud:2018,Kurashvili-Kouz-Stud:2017}), $\mu_{ij}$ (where $i,j=1,2$) are magnetic moments in the mass basis, see section \ref{Intro}. We note also that the condition (\ref{5-Anti-Rez-Flav}) is realized at a neutrino energy twice as large as in the formula (\ref{4-Anti-Rez}).

It should also be noted that accounting for neutrino mixing leads to a decrease in the effective IMM value of the muon neutrino, i.e. $\mu_{\mu\mu}^{\mathrm{ind}}=-\mu^{\mathrm{ind}}(1-\sin2\theta)$. In addition, the effect of disappearance of oscillations in the case of a muon neutrino is not realized, since the IMM and diagonal magnetic moment of the muon neutrino have the same signs.

Let us also discuss if flavor transitions could modify substantially our
results from previous sections. In principle, this can be the case if the
resonance condition (\ref{2-Rez}) overlaps with other types of neutrino
oscillation resonances. Here, we should stress that we investigate
oscillations among neutrino states with different helicity but specific
flavor. The correspon\-ding transitions with the change of flavor in a magnetic
field (spin-flavor oscillations) have long been extensively studied in a
number of publications and applied to supernova condi\-tions (see, for instance,
\cite{Ahriche:2003,*Akhmedov-Fuku-JCAP:2003,*Ando-Sato-1:2003,*Ando-Sato-2:2003,*Ando-Sato-JCAP:2003,*Yoshida-Takamura-SF:2009,Gouvea-Shalgar-1:2012,*Gouvea-Shalgar-2:2013,*Kharlanov-Shustov:2019}%
). Together with purely flavor conversions in matter (the MSW effect) for the
conditions of supernovae they have been shown to appear in outer regions with
density $\rho\lesssim10^{6}\,\text{g/cm}^{3}$. Relatively recently it has been
recognized that down the radius scale, in the densities range $\rho\sim
10^{6}{-}10^{10}\,\text{g/cm}^{3}$, \emph{the collective effects} become
dominant (see, e.g. \cite{Fogli:2007,Raffelt-opport-SN:2010}; for a review see
\cite{Duan-Full-Qian:2010,Balantekin-Fuller:2013,Mirizzi-Tamborra:2016}).
And finally, as mentioned above, the spin transition effect takes place in the region $\rho\sim10^{9}{-}10^{12}\,\text{g/cm}^{3}$ with the highest density in the close vicinity of the neutrino sphere.

\section{Possible mechanisms of nonlinear feedback \label{Feedbacks}}

Nonlinear feedback can have a significant impact on the processes of resonance neutrino conversion, affecting both the survival probabilities and the adiabaticity of the transitions.  Various feedback mechanisms that can contribute to the phenomena under consideration are discussed in the current literature.

\subsection{Compensation of the neutrino-matter and neutrino-neutrino potentials}

If an electron neutrino propagates in an environment in which other electron neutrinos are present, then it is necessary to include into the consideration the potential $\mathrm{V}_{\nu_{e}}$ describing the coherent neutrino-neutrino forward scattering (neutrino self-interaction potential) \cite{Fuller-Mayle:87,*Pantaleone-N-N:92,Notz-Raffelt:88}. Let us first consider the simplest case of a homogeneous and isotropic neutrino medium \cite{Notz-Raffelt:88}, then
\begin{equation}
\mathrm{V}_{\nu_{e}}=2\sqrt{2}\,G_{\mathrm{F}}(n_{\nu_{e}}-n_{\bar{\nu}_{e}%
})=2\sqrt{2}\,G_{\mathrm{F}}\frac{\rho_{\mathrm{B}}}{m_{\mathrm{N}}}Y_{\nu
_{e}}, \label{6-Vnu}%
\end{equation}
where $n_{\nu_{e}}$ and $n_{\bar{\nu}_{e}}$ are neutrino and antineutrino number densities, $Y_{\nu_{e}}=(n_{\nu_{e}}-n_{\bar{\nu}_{e}})/n_{\mathrm{B}}$ is the value of the neutrino fraction. The expression (\ref{6-Vnu}) for the neutrino potential should be introduced into equations (\ref{2-Hmed}), (\ref{3-Hmed-Maj}), and (\ref{2-Ve}) using the substitution $\mathrm{V}_{e}\rightarrow\mathrm{V}_{e}+\mathrm{V}_{\nu_{e}}$. As a result, from equation (\ref{2-Ve}) we obtain the following resonance condition for spin oscillations:
\begin{equation}
Y_{e}+\frac{4}{3}\,\frac{n_{\nu_{e}}-n_{\bar{\nu}_{e}}}{n_{\mathrm{B}}}%
=Y_{e}+\frac{4}{3}Y_{\nu_{e}}=\frac{1}{3}. \label{6-Rez-Ynu}%
\end{equation}
If the neutrino contribution to the full Hamiltonian (see (\ref{2-Schroed-like})) is relatively small (as it was assumed above), then (\ref{6-Rez-Ynu}) transforms to the usual resonance condition (\ref{2-Rez}), i.e. $Y_{e}\approx1/3$.

The feedback works as follows. Suppose that the resonance condition (\ref{6-Rez-Ynu}) is satisfied at some initial moment. As a result, left-handed neutrinos begin to convert to right-handed states. If they are \emph{Majorana particles}, then such transitions are equivalent to $\nu_{e}\rightarrow\bar{\nu}_{e}$ transitions. Therefore, the antineutrino number density $n_{\bar{\nu}_{e}}$ begins to grow. When neutrinos pass through the resonance region, the value of $Y_{e}$ increases (see section \ref{IMMonly}). If the variation rates of $Y_{e}$ and $n_{\bar{\nu}_{e}} $ are close, then after some time the resonance (\ref{6-Rez-Ynu}) will appear again. As a result, if $Y_{e}$ increases sufficiently slowly, then the resonance condition can be fulfilled several times during the entire transition. In this case, the condition $\mathrm{V}_{e}+\mathrm{V}_{\nu_{e}}\approx0$ holds during the entire transition, i.e., there will be an almost complete compensation of the neutrino-matter and neutrino-neutrino potentials.

In addition, we note that the adiabaticity condition now has the form (compare with (\ref{2-Ad-gen}), see also section \ref{IMMonlyMN})
\begin{equation}
\varkappa=\frac{4(2\mu_{\nu}^{\mathrm{ind}}\gamma^{-1}B_{\bot})^{2}%
}{\left\vert d\mathrm{V}_{e}/dr+d\mathrm{V}_{\nu_{e}}/dr\right\vert }\gtrsim1.
\label{6-Adia}%
\end{equation}
Derivatives in the denominator of the expression (\ref{6-Adia}), averaged over the entire transition length, can almost completely compensate each other. This will lead to a significant increase in adiabaticity during the entire transition. As a result, neutrinos with even larger energy values (as compared to those considered in sections \ref{IMMonlyMN} and \ref{Mixing}) will be involved in the conversion $\nu_{e}\rightarrow\bar{\nu}_{e}$.

From the physical point of view, the feedback mechanism considered here is analogous to the matter-neutrino resonance (MNR) phenomenon, which can occur, as expected, during flavor conversion of neutrinos produced in compact astrophysical object mergers \cite{Malkus-MNR:2014,*Malkus-McLaughlin:2016,*Vaananen-MNR:2016,*Wu-Duan:2016,*Zhu-Perego-MNR:2016,*Shalgar-MNR:2018,*Vlasenko-MNR:2018}.

The considered feedback mechanism is also manifested in neutrino-antineutrino transitions, which are realized near astrophysical objects due to the phenomenon of \emph{spin coherence} \cite{Vlasenko-Full-Cirigl:2014,*Kart-Raffelt-Propag:2015,Cirigl-Fuller-Vlas:2015}. These transitions occur with a helicity flip, and for Majorana neutrinos they are equivalent to $\nu\leftrightarrows\bar{\nu}$ transitions. However, here the helicity is flipped not due to the interaction with a medium polarized by a magnetic field, but due to the interaction with a \emph{moving medium} (with flows of matter particles and neutrino fluxes), when the medium velocity has a component perpendicular to the direction of the neutrino propagation (on this item, see also section \ref{Intro}).

The above-mentioned phenomenon has been analyzed in detail in a number of papers (including feedback), in particular, in \cite{Vlasenko-Fuller-Prospects:2014,Tian-SN:2017} (in the case of a collapsing supernova), and also in \cite{Chatelain-Volpe-Hel-cog:2017} (in the case of neutron star mergers).

If in expression (\ref{6-Vnu}) one takes into account the motion of neutrinos of the medium (see, e.g., \cite{Qian-Fuller-1-Nu-eff:95,*Fuller-Qian-nn-pot:2006}) then the conversion $\nu\rightarrow\bar{\nu}$ will occur both due to interaction with the polarized medium, and due to  interaction with the moving medium. The resonance conditions for these processes turned out to be close, and the resonance regions may overlap. Consequently, it is necessary to consider the joint action of both mechanisms causing a flip of neutrino helicity.

\subsection{Effect of oscillations on the electron fraction $Y_{e}$}

In this feedback mechanism, an essential role is played by neutrino (antineutrino) capture reactions by free nucleons near the neutrino sphere:
\begin{equation}
\nu _{e}+n\rightarrow p+e^{-},\quad \bar{\nu}_{e}+p\rightarrow n+e^{+}.
\label{6-nu-anti-nu-capt}
\end{equation}
These processes provide the heating of the matter over the neutrino sphere and cause the neutrino-driven wind \cite{Janka-Lang:2007,*Burrows-RMP:2013,Arcones-Wind:2013}.

In addition, the reactions (\ref{6-nu-anti-nu-capt}) (as well as the reverse processes) set the value of $Y_{e} $ and, therefore, determine the conditions for nucleosynthesis in neutrino heated outflows \cite{Qian-Woosley-nucleosynt:93,*Qian-Woosley-wind:96,Fuller-Meyer-n-synt:95,Arcones-Wind:2013,Balantekin-Yuksel:2002}.

Neutrino oscillations of different types (flavor and spin oscillations) can affect the rates of the reactions (\ref{6-nu-anti-nu-capt}) (see, for example, \cite{Qian-Woosley-nucleosynt:93,*Qian-Woosley-wind:96}), and therefore they can change the value of $Y_{e}$. And this, in turn, can affect the dynamics and nucleosynthesis of supernovae. On the other hand, a change in the $Y_{e}$ profile can have a backward effect on the oscillations, in particular, on the adiabaticity of transitions.

This type of feedback is most often considered in the context of flavor oscillations involving sterile neutrinos \cite{Nunokawa-Peltoniemi-ster:97,McLaughlin-Bal:99,*Fetter-Bal-ster:2003,Beun-McLaughlin:2006,Tamborra-Raffelt:2012,Wu-Fischer-ster:2014,Pllumbi-Tamborra:2015,*Xiong-Wu-Qian:2019}. In particular, in papers \cite{Wu-Fischer-ster:2014,Pllumbi-Tamborra:2015,*Xiong-Wu-Qian:2019} it was shown that taking this feedback into account can lead to a decrease in the derivative $dY_{e}/dr$ in the resonance point, and this leads to a significant increase in adiabaticity for transitions between active and sterile neutrinos.

The influence of the considered type of feedback on neutrino spin oscillations remains an open question, both in the case of transitions due to spin coherence \cite{Vlasenko-Fuller-Prospects:2014,Tian-SN:2017}, and in our case of spin oscillations in a polarized medium. Conducting the relevant research is the task of the near future.

\section{Conclusion}

In the present work we have considered helicity transitions (spin oscillations) of massive neutrinos interacting with matter polarized by an external magnetic field. Conditions are determined under which these transitions will have a resonant character in collapsing supernova matter.

For Dirac neutrinos, the transitions $\nu_{e}^{s=-1}\rightarrow\nu_{e}^{s=+1}$ are at the same time transitions to sterile states of right-handed neutrinos. In this regard, we predict the attenuation of the low-energy part of the neutrino signal from the supernova (for $\nu_{e}$ with energies $\lesssim1 $~MeV) for some limited time after the core bounce.

In the case where the Dirac neutrino has a sufficiently large magnetic moment, the ``direct'' interaction of the neutrino magnetic moment with the magnetic field can cancel the influence of the polarized medium under certain conditions. As a result, there will be an effect of oscillations disappearance, accompanied by a sharp increase of the survival probability without helicity change in a certain energy range of neutrinos. Observation of the characteristic maximum in the spectrum of electron neutrinos from supernova at relatively low energies $E\lesssim1$~MeV may indicate that neutrinos possesses a magnetic moment with a magnitude of at least $\mu_{\nu}\gtrsim10^{-17}\mu_{\mathrm{B}}$ for the conditions considered above.

Interaction with the polarized matter leads also to spin oscillations of the Majorana neutrinos. For Majorana neutrinos, the transitions $\nu_{e}^{s=-1}\rightarrow\nu_{e}^{s=+1}$ are at the same time the neutrino-antineutrino transitions. This phenomenon will lead to distortion of the low-energy part of the neutrino signal from the supernova.

Accounting for neutrino mixing (in the two-flavor approximation) can significantly increase the magnitude of the effective induced magnetic moment (IMM) of the electron neutrino in the medium and, as a consequence, increase the adiabaticity of transitions. This will lead to the fact that neutrinos with even higher energy values (up to ${\sim}2$ MeV for Dirac and up to ${\sim}4$ MeV for Majorana neutrino under the conditions discussed in sections \ref{IMMonly} and \ref{IMMonlyMN}) will be involved into the transitions with the helicity flip. Along with this, the mixing will lead to a decrease in the effective value of IMM (and to a decrease in adiabaticity) for the muon neutrino. In addition, in the case of a muon neutrino, there will be no effect of oscillations disappearance as a result of the compensation of the interaction of the intrinsic magnetic moment with a magnetic field and the influence of a polarized medium.

Section \ref{Feedbacks} provides a brief overview of the feedback mechanisms that accompany the neutrino conversion processes and can lead, in particular, to an additional increase in the adiabaticity of neutrino transitions with a helicity flip.

In conclusion, we note that a detailed observation of the low-energy part of the neutrino signal from supernovae will be possible with new neutrino detectors which employ the phenomenon of neutrino coherent scattering by nuclei and which have a low threshold energy (of the order of several keV or even lower) \cite{Scholberg-SN-Detection:2012,*Biassoni:2012,*Gallo-Rosso-PoS:2018,*Kozynets:2019,Mirizzi-Tamborra:2016}.

\section*{Acknowledgements}

The authors are grateful to P.A. Eminov and A.E. Lobanov for useful discussions of a number of issues addressed in this article. The authors also thank the editor W.C. Haxton for careful consideration of the manuscript. The work of A.G. and A.T. was supported by Russian Ministry of Science and Education, project \#3.9911.2017/BasePart, and also by the Russian Foundation for Basic Research under grants No. 16-02-01023-a and No. 17-52-53133-GFEN.


\bibliography{Grig-Kup-Ter-0407-PLB}

\end{document}